\documentclass[12pt]{article}
\usepackage{epsfig}
\usepackage{subfigure}
\usepackage{graphicx}
\usepackage{amssymb}
\usepackage{amsfonts}
\usepackage{amsmath}
\usepackage{cite}
\usepackage{multirow}

\begin{document}

\begin{titlepage}
\begin{center}
\hfill    IFIC/16-68

\vskip 1cm

{\large \bf {Vector-like Quarks at the Origin of Light Quark Masses and Mixing}}

\vskip 1cm

Francisco J. Botella  $^{a}$ \footnote{Francisco.J.Botella@uv.es}, 
G. C. Branco  $^{b}$ \footnote{gbranco@tecnico.ulisboa.pt}, 
Miguel Nebot $^{b}$ \footnote{nebot@cftp.tecnico.ulisboa.pt},
M. N. Rebelo $^{b}$ \footnote{rebelo@tecnico.ulisboa.pt} \\
and J. I. Silva-Marcos $^b$ \footnote{juca@cftp.tecnico.ulisboa.pt},

\vspace{1.0cm}

{\it $^a$ Departament de F\' \i sica Te\`orica and IFIC,
Universitat de Val\`encia-CSIC, \\
E-46100 Burjassot, Spain.} \\
{\it $^b$ Departamento de F\'\i sica and Centro de F\' \i sica Te\' orica
de Part\' \i culas (CFTP),
Instituto Superior T\' ecnico (IST), Universidade de Lisboa (UL), Av. Rovisco Pais, P-1049-001 
Lisboa, Portugal. }

\end{center}

\vskip 3cm

\begin{abstract}
We show how a novel fine-tuning problem present in the Standard Model can
be solved through the introduction of a single flavour symmetry G, together with 
three $Q = - 1/3$ quarks, three $Q = 2/3$ quarks, as well as a complex singlet scalar.
The symmetry G is extended to the additional fields and it is an exact symmetry of the 
Lagrangian, only spontaneously broken by the vacuum. Specific examples are given 
and a phenomenological analysis of the main features of the model is presented. 
It is shown that even for vector-like quarks with masses accessible at
the LHC, one can have realistic quark masses and mixing, while respecting the
strict constraints on process arising from flavour changing neutral 
currents (FCNC). The vector-like quark decay channels are also described. 
\end{abstract}

\end{titlepage}

\newpage

\section{Introduction}

In the framework of the Standard Model (SM), the Brout-Englert-Higgs
mechanism is responsible not only for the breaking of the gauge symmetry but
also for the generation of fermion masses, through the Yukawa interactions
of the scalar doublet with quarks and leptons. Understanding the observed
pattern of fermion masses and mixing remains a fundamental open question in
Particle Physics. On the experimental side, there has been great progress
and at present the moduli of the Cabibbo Kobayashi Maskawa (CKM) matrix, $%
V_{CKM}$, are reasonably well known \cite{Agashe:2014kda, Amhis:2014hma,
Porter:2016mps} with clear evidence that the mixing matrix is non-trivially
complex, even if one allows for the presence of New Physics (NP) beyond the
SM \cite{Botella:2005fc, Charles:2004jd, Bona:2005vz}. 
The discovery of the Higgs particle at LHC \cite%
{Aad:2012tfa, Chatrchyan:2012xdj} renders specially important the
measurement of the Higgs couplings to quarks and charged leptons. The strength of
these couplings is fixed within the SM, but at present we have a poor
knowledge of their actual value.

Recently, it has been pointed out \cite{Botella:2016krk} that there is a
novel fine-tuning problem in the SM. Contrary to conventional wisdom, in the
SM without extra flavour symmetries, quark mixing is naturally large, in
spite of the large quark mass hierarchy. In fact, even in the extreme chiral
limit, where only the third family acquires mass, mixing is meaningful, and
in general of order one. It is possible to solve this fine-tuning problem
through the introduction of a simple flavour symmetry which leads to $%
V_{CKM} = {1\>\!\!\!\mathrm{I}}$. The challenge is then to achieve quark
mixing and masses for the first two generations.

In this paper we face this challenge and put forward a framework where a
flavour symmetry leads to $V_{CKM} = {1\>\!\!\!\mathrm{I}}$ in leading
order, with light quark masses and mixing being generated by the presence of
vector-like quarks (VlQ). We introduce three up-type VlQ and three down-type
VlQ and derive effective mass squared Hermitian mass matrices for the
standard quarks. In this framework one finds a natural explanation why $%
|V_{13}|^2 + |V_{23}|^2$ is very small, while allowing for an adequate
Cabibbo mixing. In this model there are Z-mediated and Higgs-mediated
Flavour Changing Neutral Currents (FCNC) which are naturally suppressed. The
VlQs can have masses of the order of one TeV which are at the reach of the
second run of LHC \cite{Aad:2015tba, Aad:2015mba, Aad:2015kqa, Aad:2015voa,
Aad:2016qpo, Aad:2016shx, Chatrchyan:2013uxa, Khachatryan:2015axa,
Khachatryan:2015gza, Khachatryan:2015oba}. At this stage, it is worth
emphasising that VlQs have been extensively studied in the literature 
\cite{delAguila:1985ne, Branco:1986my,
delAguila:1987nn,delAguila:1987se, Langacker:1988ur, delAguila:1989rq,
Bento:1991ez, Cheng:1991rr, Lavoura:1992qd, Branco:1992wr, Barger:1995dd,
Barenboim:1997qx, Barenboim:2000zz,  Barenboim:2001fd, AguilarSaavedra:2002kr,
AguilarSaavedra:2004mt, Botella:2008qm, Higuchi:2009dp, Cacciapaglia:2010vn,
Botella:2012ju, Aguilar-Saavedra:2013qpa, Buchkremer:2013bha,
Fajfer:2013wca, Ellis:2014dza, Alok:2015iha, Cacciapaglia:2015vrx,
Ishiwata:2015cga, Bobeth:2016llm, Buchkremer:2012dn} and arise in a variety
of frameworks, including E6 GUTS, models with extra-dimensions, models
providing solutions to the strong CP problem without axions, etc. It has
also been pointed out \cite{Dermisek:2012ke} that non-supersymmetric
extensions of the SM with VlQs can achieve unification of gauge couplings.

The paper is organised as follows: in the next section we briefly review a
novel fine-tuning problem of the Standard Model and illustrate how it can be
solved through the addition of a flavour symmetry. In section 3, we show how
a realistic quark mass spectrum and pattern of mixing can be generated
through the introduction of vector-like quarks and a complex scalar singlet.
In section 4 we give specific examples examining the constraints arising
from various FCNC processes and describing the vector-like quark decay
channels. Finally, our conclusions are presented in section 5.

\section{Hierarchy and Alignment through a Flavour Symmetry}

\subsection{A Novel Fine-Tuning Problem in the Standard Model}

Recently, it has been pointed out \cite{Botella:2016krk} that in the SM
there is a novel fine-tuning problem which stems from the fact that the
natural value of $|V_{13}|^2 + |V_{23}|^2$ is of order one in the SM, to be
compared to the experimental value of $1.6 \times 10^{-3}$. In order to
obtain this result, one can examine the extreme chiral limit where only the
third family of quarks has mass while all other quarks are massless. In this
limit, the general quark mass matrices can be written 
\begin{equation}
M_{d}={U_{L}^{d}}^{\dagger }\ \mbox{diag}(0,0,m_{b})\ {U_{R}^{d}},\qquad
M_{u}={U_{L}^{u}}^{\dagger }\ \mbox{diag}(0,0,m_{t})\ {U_{R}^{u}}
\label{loo}
\end{equation}
where $U_{L,R}^{d,u}$ are arbitrary unitary matrices. The quark mixing
matrix is then $V^{0}={U_{L}^{u}}^{\dagger }{U_{L}^{d}}$. Using the freedom
to redefine the quark fields of the massless quarks through two-by-two
unitary matrices one can show that the quark mixing can be described by an
orthogonal two-by-two rotation connecting only the ($c$, $t$) and ($s$, $b$)
quarks. This angle is arbitrary and expected to be of order one.

\subsection{A Flavour Symmetry Leading to Small Mixing}

Let us consider the SM and introduce the following symmetry: 
\begin{equation}
\begin{array}{lllll}
Q_{L1}^{0}\rightarrow e^{i\tau }\ Q_{L1}^{0} &  & Q_{L2}^{0}\rightarrow
e^{-2i\tau }\ Q_{L2}^{0} &  & Q_{L3}^{0}\rightarrow e^{-i\tau }\ Q_{L3}^{0}
\\ 
&  &  &  &  \\ 
d_{R1}^{0}\rightarrow e^{-i\tau }d_{R1}^{0} &  & d_{R2}^{0}\rightarrow
e^{-i\tau }d_{R2}^{0} &  & d_{R3}^{0}\rightarrow e^{-2i\tau }d_{R3}^{0} \\ 
&  &  &  &  \\ 
u_{R1}^{0}\rightarrow e^{i\tau }u_{R1}^{0} &  & u_{R2}^{0}\rightarrow
e^{i\tau }u_{R3}^{0} &  & u_{R3}^{0}\rightarrow u_{R3}^{0}\ \quad ;\quad
\Phi \rightarrow e^{i\tau }\Phi%
\end{array}
\label{SSM}
\end{equation}%
where the $Q_{Lj}^{0}$ are left-handed quark doublets, $d_{Rj}^{0}$ and $%
u_{Rj}^{0}$ are right-handed quark singlets and $\Phi $ denotes the Higgs
doublet. The Yukawa interactions are given by: 
\begin{equation}
\mathcal{L}_{\mathrm{Y}}=\left[ -{\overline{Q}_{Li}^{0}}\ \,\Phi \,Y_{d}\,\
d_{Rj}^{0}-\,{\overline{Q}_{Li}^{0}}\ \tilde{\Phi}\,Y_{u}\,\ u_{Rj}^{0}%
\right] +\mbox{h.c.},
\end{equation}%
and this symmetry leads to the following pattern of texture zeros for the
Yukawa couplings: 
\begin{equation}
Y_{d}=\left[ 
\begin{array}{ccc}
0 & 0 & 0 \\ 
0 & 0 & 0 \\ 
0 & 0 & \times%
\end{array}%
\right] ,\qquad Y_{u}=\left[ 
\begin{array}{ccc}
0 & 0 & 0 \\ 
0 & 0 & 0 \\ 
0 & 0 & \times%
\end{array}%
\right]  \label{ydyu}
\end{equation}%
which lead to $V_{CKM}$ equal to the identity with only one non-zero quark
mass in each charge sector.

\section{Vector-like quarks and generation of realistic quark masses and
mixing}

A possible mechanism to generate masses for the light standard-like quarks
is to introduce vector-like quarks. In our examples we particularise for the
case of three down ($D^0_{Li}$, $D^0_{Ri}$,) and three up ($U^0_{Li}$, $%
U^0_{Ri}$,) vector-like isosinglet quarks. With the introduction of these
additional isosinglet quarks the Yukawa interactions can now be denoted as: 
\begin{equation}
\mathcal{L}_{\mathrm{Y}}=\left[- {\overline Q_{Li}^{0}} \,\Phi\, (Y_d)_{i
\alpha} \, d^0_{R\alpha} - \, {\overline Q_{Li}^{0}}\tilde\Phi \, (Y_u)_{i
\beta} \, u^0_{R\beta} \right] + \mbox{h.c.},  \label{YYY}
\end{equation}
here the index $i$ runs from 1 to 3, as in the SM, while the indices $\alpha$
and $\beta$ cover all right-handed quark singlets of the down and up sector,
respectively. The following generic bare mass terms must also be introduced
in the Lagrangian: 
\begin{equation}
\mathcal{L}_{b.m.} = [- {\ \overline D^0_{Lj}} (\eta_d)_{j \alpha} \,
d^0_{R\alpha} - {\overline U^0_{Lk}} (\eta_u)_{k \beta} \, u^0_{R\beta} ] + %
\mbox{h.c.}  \label{bbb}
\end{equation}
here the indices $j$ and $k$ run over all left-handed vectorial quarks in
each sector. As mentioned before, in all examples that follow $i, j$ and $k$
run from 1 to 3 and therefore $\alpha$ and $\beta$ run from 1 to 6
(obviously $D^0_{Ri} \equiv d^0_{Ri+3}$ and $U^0_{Ri} \equiv u^0_{Ri+3}$). In
what follows we extend the discrete flavour symmetry introduced in
subsection 2.2 and we introduce a complex scalar singlet $S$. This scalar
singlet will couple to the quark singlets in the following way: 
\begin{equation}
\mathcal{L}_{\mathrm{g}} = [- {\ \overline D^0_{Lj}} [({g_d})_{j \alpha} S +
({g_d^\prime})_{j \alpha} S^\ast ] \, d^0_{R\alpha} - {\overline U^0_{Lk}} [(%
{g_u})_{k \beta} S + ({g_u^\prime})_{k \beta} S^\ast ] \, u^0_{R\beta} ] + %
\mbox{h.c.}  \label{ggg}
\end{equation}
We assume that the modulus of the vacuum expectation value of the field $S$
is of an order of magnitude higher than the electroweak scale. After
spontaneous symmetry breaking the following mass terms are generated: 
\begin{equation}
\mathcal{L}_{\mathrm{M}} = \left[- \frac{v}{\sqrt{2}}\, {\ \overline d^0_{Li}%
} (Y_d)_{i \alpha} \, d^0_{R\alpha} - \frac{v}{\sqrt{2}} {\overline u^0_{Li}}
(Y_u)_{i \alpha} \, u^0_{R\alpha} - {\overline D^0_{Li}} (\mu_d)_{i \alpha}
\, d^0_{R\alpha} - {\overline U^0_{Li}} (\mu_u)_{i \alpha} \, u^0_{R\alpha} %
\right] + \mbox{h.c.}
\end{equation}
These terms can be written in a more compact form, as: 
\begin{equation}
\mathcal{L}_{\mathrm{M}} = - \left( {\ \overline d^0_{L}} \ {\overline
D^0_{L}} \right) \mathcal{M}_d \, \left( 
\begin{array}{c}
d^0_{R} \\ 
D^0_{R}%
\end{array}
\right) - \left( {\ \overline u^0_{L}} \ {\overline U^0_{L}} \right) 
\mathcal{M}_u \, \left( 
\begin{array}{c}
u^0_{R} \\ 
U^0_{R}%
\end{array}
\right)
\end{equation}
with $6 \times 6$ mass matrices, $\mathcal{M}_d$ and $\mathcal{M}_u$,
denoted as: 
\begin{equation}
\mathcal{M}_d = \left( 
\begin{array}{cc}
m_d & \omega_d \\ 
X_d & M_d%
\end{array}
\right) \quad \mathcal{M}_u = \left( 
\begin{array}{cc}
m_u & \omega_u \\ 
X_u & M_u%
\end{array}
\right)  \label{not}
\end{equation}

\subsection{ Structure of Charged and Neutral currents}

The matrices $\mathcal{M}_{d}$ and $\mathcal{M}_{u}$ will be diagonalised
through the following unitary transformations: 
\begin{equation}
\begin{array}{lll}
\left( 
\begin{array}{c}
d_{L}^{0} \\ 
D_{L}^{0}%
\end{array}%
\right) =\left( 
\begin{array}{c}
A_{dL} \\ 
B_{dL}%
\end{array}%
\right) \left( 
\begin{array}{c}
d_{L}%
\end{array}%
\right) \equiv \mathcal{U}_{L}^{d}d_{L} &  & \left( 
\begin{array}{c}
u_{L}^{0} \\ 
U_{L}^{0}%
\end{array}%
\right) =\left( 
\begin{array}{c}
A_{uL} \\ 
B_{uL}%
\end{array}%
\right) \left( 
\begin{array}{c}
u_{L}%
\end{array}%
\right) \equiv \mathcal{U}_{L}^{u}u_{L} \\ 
&  &  \\ 
\left( 
\begin{array}{c}
d_{R}^{0} \\ 
D_{R}^{0}%
\end{array}%
\right) \equiv \mathcal{U}_{R}^{d}d_{R} &  & \left( 
\begin{array}{c}
u_{R}^{0} \\ 
U_{R}^{0}%
\end{array}%
\right) \equiv \mathcal{U}_{R}^{u}u_{R}%
\end{array}%
\end{equation}%
where $A_{dL}$, $B_{dL}$, $A_{uL}$ and $B_{uL}$ are $3\times 6$ matrices;
the matrices $\mathcal{U}_{L}^{d}$, $\mathcal{U}_{R}^{d}$, $\mathcal{U}%
_{L}^{u}$ and $\mathcal{U}_{R}^{u}$ are unitary $6\times 6$ matrices and $%
(d_{L})$, $(d_{R})$, $(u_{L})$ and $(u_{R})$ stand for the components of the
six down and six up mass eigenstate quarks. Unitarity implies that 
\begin{equation}
\begin{array}{l}
\left( 
\begin{array}{c}
A_{dL} \\ 
B_{dL}%
\end{array}%
\right) \left( A_{dL}^{\dagger }\ \ B_{dL}^{\dagger }\right) =\left( 
\begin{array}{c}
A_{dL}A_{dL}^{\dagger }\ \ A_{dL}B_{dL}^{\dagger } \\ 
B_{dL}A_{dL}^{\dagger }\ \ B_{dL}B_{dL}^{\dagger }%
\end{array}%
\right) =\left( 
\begin{array}{cc}
{1\>\!\!\!\mathrm{I}}_{3\times 3} & 0 \\ 
0 & {1\>\!\!\!\mathrm{I}}_{3\times 3}%
\end{array}%
\right)  \\ 
\\ 
\left( A_{dL}^{\dagger }B_{dL}^{\dagger }\right) \left( 
\begin{array}{c}
A_{dL} \\ 
B_{dL}%
\end{array}%
\right) =A_{dL}^{\dagger }A_{dL}+B_{dL}^{\dagger }B_{dL}={1\>\!\!\!\mathrm{I}%
}_{6\times 6}%
\end{array}%
\end{equation}%
with similar relations for the up-type matrices. The charged currents are
given by: 
\begin{equation}
\mathcal{L}_{\mathrm{W}}=-\frac{g}{\sqrt{2}}\left( {\overline{u}_{L}^{0}}%
\gamma ^{\mu }d_{L}^{0}\right) \mathbf{W}_{\mu }^{+}+\mbox{h.c.}=-\frac{g}{%
\sqrt{2}}\left( {\overline{u}_{L}}V\gamma ^{\mu }d_{L}\right) \mathbf{W}%
_{\mu }^{+}+\mbox{h.c.}
\end{equation}%
with $V=A_{uL}^{\dagger }A_{dL}$. The couplings of the $Z$ boson are of the
form: 
\begin{equation}
\mathcal{L}_{\mathrm{Z}}=\frac{g}{\cos {\theta _{W}}}Z_{\mu }\left[ \frac{1}{%
2}\left( {\overline{u}_{L}}W^{u}\gamma ^{\mu }u_{L}-{\overline{d}_{L}}%
W^{d}\gamma ^{\mu }u_{L}\right) -\sin ^{2}\theta _{W}\left( \frac{2}{3}%
\overline{u}\gamma ^{\mu }u-\frac{1}{3}\overline{d}\gamma ^{\mu }d\right) %
\right] 
\end{equation}%
with $W^{d}=V^{\dagger }V$ and $W^{u}=VV^{\dagger }$. After spontaneous
symmetry breaking, Eqs.~(\ref{YYY}), (\ref{bbb}) and (\ref{ggg}) give rise
to the physical quark masses, which together with the couplings to the
SM-like Higgs can be denoted as: 
\begin{equation}
\begin{array}{l}
\mathcal{L}_{\mathrm{Mh}}=-\overline{d}_{L}\mathcal{D}_{d}d_{R}-\overline{u}%
_{L}\mathcal{D}_{u}u_{R}-\frac{\sqrt{2}G^{+}}{v}\left[ \overline{u}_{L}V%
\mathcal{D}_{d}d_{R}-\overline{u}_{R}\mathcal{D}_{u}Vd_{L}\right]  \\ 
\\ 
-i\frac{G^{0}}{v}\left[ \overline{d}_{L}W^{d}\mathcal{D}_{d}d_{R}-\overline{%
u}_{L}W^{u}\mathcal{D}_{u}u_{R}\right] -\frac{h}{v}\left[ \overline{d}%
_{L}W^{d}\mathcal{D}_{d}d_{R}+\overline{u}_{L}W^{u}\mathcal{D}_{u}u_{R}%
\right] +\mbox{h.c.}%
\end{array}%
\end{equation}%
where $h$ is the Higgs field, $v$ is the vacuum expectation value of the
neutral component of the Higgs doublet and $G^{+}$ and $G^{0}$ are the
would-be Goldstone bosons, $\mathcal{D}_{d}$ and $\mathcal{D}_{u}$ are the
six by six diagonal quark mass matrices.

\subsection{Extension of the symmetry to the full Lagrangian}

In the fermion sector, as mentioned above, we introduce three down-type and
three up-type vector-like quarks. In the scalar sector, in addition to the
standard Higgs, we introduce a complex scalar $S$. We extend the symmetry to
the full Lagrangian, with the new fields transforming in the following way
under the family symmetry:%
\begin{equation}
\begin{array}{lllll}
D_{L1}^{0}\rightarrow e^{-3i\tau }\ D_{L1}^{0} &  & D_{L2}^{0}\rightarrow
e^{-2i\tau }\ D_{L2}^{0} &  & D_{L3}^{0}\rightarrow e^{-i\tau }\ D_{L3}^{0}
\\ 
&  &  &  &  \\ 
D_{R1}^{0}\rightarrow e^{-2i\tau }\ D_{R1}^{0} &  & D_{R2}^{0}\rightarrow
e^{-3i\tau }\ D_{R2}^{0} &  & D_{R3}^{0}\rightarrow D_{R3}^{0} \\ 
&  &  &  &  \\ 
U_{L1}^{0}\rightarrow e^{-i\tau }\ U_{L1}^{0} &  & U_{L2}^{0}\rightarrow
U_{L2}^{0} &  & U_{L3}^{0}\rightarrow e^{i\tau }\ U_{L3}^{0} \\ 
&  &  &  &  \\ 
U_{R1}^{0}\rightarrow U_{R1}^{0} &  & U_{R2}^{0}\rightarrow e^{-i\tau }\
U_{R2}^{0} &  & U_{R3}^{0}\rightarrow e^{2i\tau }\ U_{R3}^{0}\ \quad ;\quad
S\rightarrow e^{i\tau }\ S%
\end{array}%
\end{equation}%
together with the transformations for the standard-like quarks specified in
Eq.~(\ref{SSM}). The singlet scalar S is introduced in order to be able to
obtain realistic quark masses and mixing, without breaking the symmetry at
the Lagrangian level. An alternative option would be to softly break the
symmetry through the introduction of bare mass terms for the heavy fermions.
Since the symmetry would be only softly broken, the model would maintain its
renormalizability. We find it more appealing to have the symmetry exact, at
the Lagrangian level. We assume that the scale of spontaneous symmetry
breaking of the $S$ fields is higher than the electroweak scale.

Table 1 and Table 2 summarise the information on the combination of the
different fermionic charges and allow to infer what is the pattern of the
mass matrices. The first three rows come from Yukawa terms of the form given
by Eq.~(\ref{YYY}) and therefore are only allowed when the fermionic charge
cancels the one coming from the scalar doublet. In these cases we write this
charge explicitly. In the forbidden terms we put a bullet sign. The last
three rows come from bare mass terms of the form given by Eq.~(\ref{bbb}) or
else from couplings to the field $S$. We denote with 1 the entries
corresponding to allowed bare mass terms and by the fermionic charges those
terms that allow coupling to either $S$ or $S^*$. Again we use bullets for
the forbidden terms. The introduction of these singlet scalar field provides
a rationale for the choice of terms that would otherwise softly break the
symmetry and would look arbitrary.

\begin{table}[htb]
\caption{Down sector, summary of transformation properties}
\label{Table:down}
\begin{center}
\begin{tabular}{|c|c|c|c|c|c|c|}
\cline{2-7}
\multicolumn{1}{c|}{} & $\left( 
\begin{array}{c}
d_{R1}^{0}\\ 
-\tau%
\end{array}
\right) $ & $\left( 
\begin{array}{c}
d_{R2}^{0}\\ 
-\tau%
\end{array}
\right) $ & $\left( 
\begin{array}{c}
d_{R3}^{0} \\ 
-2 \tau%
\end{array}
\right) $ & $\left( 
\begin{array}{c}
D_{R1}^{0} \\ 
- 2 \tau%
\end{array}
\right) $ & $\left( 
\begin{array}{c}
D_{R2}^{0} \\ 
- 3 \tau%
\end{array}
\right) $ & $\left( 
\begin{array}{c}
D_{R3}^{0} \\ 
0%
\end{array}
\right) $ \\ \hline
$\left( 
\begin{array}{c}
{\overline Q_{L1}^{0}} \\ 
-\tau%
\end{array}
\right) $ & $\bullet$ & $\bullet$ & $\bullet$ & $\bullet$ & $\bullet$ & $%
-\tau $ \\ \hline
$\left( 
\begin{array}{c}
{\overline Q_{L2}^{0}} \\ 
2 \tau%
\end{array}
\right) $ & $\bullet$ & $\bullet$ & $\bullet$ & $\bullet$ & $-\tau $ & $%
\bullet$ \\ \hline
$\left( 
\begin{array}{c}
{\overline Q_{L3}^{0}} \\ 
\tau%
\end{array}
\right) $ & $\bullet$ & $\bullet$ & $-\tau $ & $-\tau $ & $\bullet$ & $%
\bullet$ \\ \hline
$\left( 
\begin{array}{c}
{\overline D_{L1}^{0}} \\ 
3 \tau%
\end{array}
\right) $ & $\bullet$ & $\bullet$ & $\tau$ & $\tau$ & $1$ & $\bullet$ \\ 
\hline
$\left( 
\begin{array}{c}
{\overline D_{L2}^{0}} \\ 
2 \tau%
\end{array}
\right) $ & $\tau $ & $\tau $ & $1$ & $1$ & $-\tau $ & $\bullet$ \\ \hline
$\left( 
\begin{array}{c}
{\overline D_{L3}^{0}} \\ 
\tau%
\end{array}
\right) $ & $1 $ & $1 $ & $-\tau $ & $-\tau $ & $\bullet $ & $\tau $ \\ 
\hline\hline
\end{tabular}%
\end{center}
\end{table}

\begin{table}[htb]
\caption{Up sector, summary of transformation properties}
\label{Table:up}
\begin{center}
\begin{tabular}{|c|c|c|c|c|c|c|}
\cline{2-7}
\multicolumn{1}{c|}{} & $\left( 
\begin{array}{c}
u_{R1}^{0} \\ 
\tau%
\end{array}
\right) $ & $\left( 
\begin{array}{c}
u_{R2}^{0} \\ 
\tau%
\end{array}
\right) $ & $\left( 
\begin{array}{c}
u_{R3}^{0} \\ 
0%
\end{array}
\right) $ & $\left( 
\begin{array}{c}
U_{R1}^{0} \\ 
0%
\end{array}
\right) $ & $\left( 
\begin{array}{c}
U_{R2}^{0} \\ 
- \tau%
\end{array}
\right) $ & $\left( 
\begin{array}{c}
U_{R3}^{0} \\ 
2 \tau%
\end{array}
\right) $ \\ \hline
$\left( 
\begin{array}{c}
{\overline Q_{L1}^{0}} \\ 
-\tau%
\end{array}
\right) $ & $\bullet$ & $\bullet$ & $\bullet$ & $\bullet$ & $\bullet$ & $%
\tau $ \\ \hline
$\left( 
\begin{array}{c}
{\overline Q_{L2}^{0}} \\ 
2 \tau%
\end{array}
\right) $ & $\bullet$ & $\bullet$ & $\bullet$ & $\bullet$ & $\tau $ & $%
\bullet$ \\ \hline
$\left( 
\begin{array}{c}
{\overline Q_{L3}^{0}} \\ 
\tau%
\end{array}
\right) $ & $\bullet$ & $\bullet$ & $\tau $ & $\tau $ & $\bullet$ & $\bullet$
\\ \hline
$\left( 
\begin{array}{c}
{\overline U_{L1}^{0}} \\ 
\tau%
\end{array}
\right) $ & $\bullet$ & $\bullet$ & $\tau$ & $\tau$ & $1$ & $\bullet$ \\ 
\hline
$\left( 
\begin{array}{c}
{\overline U_{L2}^{0}} \\ 
0%
\end{array}
\right) $ & $\tau $ & $\tau $ & $1$ & $1$ & $-\tau $ & $\bullet$ \\ \hline
$\left( 
\begin{array}{c}
{\overline U_{L3}^{0}} \\ 
- \tau%
\end{array}
\right) $ & $1 $ & $1 $ & $-\tau $ & $-\tau $ & $\bullet $ & $\tau $ \\ 
\hline\hline
\end{tabular}%
\end{center}
\end{table}

\subsection{Effective Hermitian squared mass matrix}

The $6 \times 6$ mass matrices $\mathcal{M}_{d}$, $\mathcal{M}_{u}$ are
diagonalised through the bi-unitary transformations: 
\begin{equation}
\mathcal{U}_{L}^{d \dagger} \mathcal{M}_{d}\ \mathcal{U}_{R}^d = \mathcal{D}%
_d \equiv \mbox{diag} (d_d, D_d)  \label{plus}
\end{equation}
where $d_d \equiv \mbox{diag}\ (m_d, m_s, m_b)$, $D_d \equiv \mbox{diag}\
(M_{D1}, M_{D2}, M_{D3})$ and with $M_{Di}$ standing for the heavy $Q = -
1/3 $ quark masses. An analogous equation holds for $\mathcal{M}_{u}$. In
our examples the diagonalisation of $\mathcal{M}_{d}$ and $\mathcal{M}_{u}$
is done through an exact numerical calculation. However, in order to have an
idea of the main physical features involved, it is useful to perform an
approximate evaluation of $\mathcal{U}_L^d$, $\mathcal{U}_L^u$ and of the
quark mass eigenvalues. For this purpose, it is useful to write $\mathcal{U}%
_L^d$, $\mathcal{U}_L^u$ in block form: 
\begin{equation}
\mathcal{U}_L = \left( 
\begin{array}{cc}
K & R \\ 
S & T%
\end{array}
\right)
\end{equation}
where $K$, $R$, $S$, $T$ are $3 \times 3$ matrices. For simplicity, we drop
the indices $d$, and $u$. In Appendix A we show that the deviations of the
unitarity of the matrix $K$ are naturally small: 
\begin{equation}
K K^\dagger = {1\>\!\!\!\mathrm{I}} - R R^\dagger  \label{23}
\end{equation}
with 
\begin{equation}
R \approx \frac{(mX^\dagger + \omega M^\dagger) T}{D^2} \approx (m/M)
\label{24}
\end{equation}
and 
\begin{equation}
K^\dagger K = {1\>\!\!\!\mathrm{I}} - S^\dagger S
\end{equation}
with 
\begin{equation}
S \approx \left( \frac{X m^\dagger + M \omega^\dagger}{X X^\dagger + M
M^\dagger} \right) K  \label{26}
\end{equation}
The matrices $K_d$, $K_u$ can be evaluated from an effective Hermitian
squared matrix $\mathcal{H}_{eff}$ through: 
\begin{equation}
K^{-1} \mathcal{H}_{eff} K = d^2
\end{equation}
with 
\begin{equation}
\mathcal{H}_{eff} = (m m^\dagger + \omega \omega^\dagger) - (mX^\dagger +
\omega M^\dagger) (X X^\dagger + M M^\dagger)^{-1} ( X m^\dagger + M
\omega^\dagger)  \label{28}
\end{equation}
The derivation of $\mathcal{H}_{eff}$ is given in Appendix A.

\section{Realistic Examples}

Following the notation in Eq.~(\ref{not}) we present one full realistic
example; mass matrices are given at the $M_{Z}$ scale in units of GeV. Among
the matrices coming from electroweak symmetry breaking we have (i) the mass
matrices that connect light (ordinary) quarks among themselves $m_{d}$ and $%
m_{u}$; they should contain the dominant contributions to the $b$ and $t$
quarks masses -allowed by the symmetry in Eq.~(\ref{SSM}) - 
\begin{equation}
m_{d}=\left( 
\begin{array}{ccc}
0 & 0 & 0 \\ 
0 & 0 & 0 \\ 
0 & 0 & 3.07015%
\end{array}%
\right) ,\ \ m_{u}=\left( 
\begin{array}{ccc}
0 & 0 & 0 \\ 
0 & 0 & 0 \\ 
0 & 0 & 185.142%
\end{array}%
\right) ,  \label{masas pequenas}
\end{equation}%
and (ii) the upper right off-diagonal blocks of the quark mass matrices that
read 
\begin{equation}
\omega _{d}=\left( 
\begin{array}{ccc}
0 & 0 & 0.061403 \\ 
0 & 0.39912 & 0 \\ 
0.39912 & 0 & 0%
\end{array}%
\right) ,\ \ \omega _{u}=\left( 
\begin{array}{ccc}
0 & 0 & 0.00925708 \\ 
0 & 3.70283 & 0 \\ 
185.142 & 0 & 0%
\end{array}%
\right) .  \label{Delta I=1/2 masses off diagonal}
\end{equation}%
The matrices in Eqs.~(\ref{masas pequenas}), (\ref{Delta I=1/2 masses off
diagonal}) are proportional to the Higgs vacuum expectation value, therefore
its matrix elements should be of the same order of magnitude or smaller than
the corresponding bottom or top mass. The heavy vectorial quarks mass
sectors $M_{d}$ and $M_{u}$ are 
\begin{equation}
M_{d}=\left( 
\begin{array}{ccc}
767.538 & 0 & 0 \\ 
0 & 1535.08 & 0 \\ 
0 & 0 & 1842.09%
\end{array}%
\right) ,\ \ M_{u}=\left( 
\begin{array}{ccc}
1295.99 & 0 & 0 \\ 
0 & 1481.13 & 0 \\ 
0 & 0 & 2221.7%
\end{array}%
\right) .  \label{masas grandes}
\end{equation}%
These matrices fix approximately the masses of the new heavy vectorial
singlet quarks. Finally, the other matrices that connect the heavy sector
with the light one are 
\begin{eqnarray}
X_{d} &=&\left( 
\begin{array}{ccc}
0 & 0 & -115.131 \\ 
-262.498i & 46.0523 & 460.523-230.261i \\ 
486.312 & 0 & 368.418%
\end{array}%
\right) ,  \label{h_l singlet mass matrices} \\
X_{u} &=&\left( 
\begin{array}{ccc}
0 & 0 & 68.4281 \\ 
-212.913 & -185.142 & 0 \\ 
416.569 & 0 & 0%
\end{array}%
\right) .  \label{h_l singlet mass matrix 2}
\end{eqnarray}%
The scale of the matrices in Eqs.~(\ref{masas grandes}), (\ref{h_l singlet
mass matrices}) and (\ref{h_l singlet mass matrix 2}) is, for most entries
similar or higher than the electroweak scale: the off-diagonal matrices
being always smaller than the heavy mass matrix sectors in order to be the
responsible of giving mass to the light quark masses and generating the CKM
mixing. Following the standard diagonalization of $\mathcal{M}_{d}$ and $%
\mathcal{M}_{u}$ the quark mass spectrum (in GeV at the $M_{Z}$ scale) is: 
\begin{equation}
\left( 
\begin{array}{c}
m_{d} \\ 
m_{s} \\ 
m_{b} \\ 
m_{D_{1}} \\ 
m_{D_{2}} \\ 
m_{D_{3}}%
\end{array}%
\right) =\left( 
\begin{array}{c}
0.0027 \\ 
0.068 \\ 
2.9 \\ 
775 \\ 
1621 \\ 
1957%
\end{array}%
\right) ,\qquad \left( 
\begin{array}{c}
m_{u} \\ 
m_{c} \\ 
m_{t} \\ 
m_{U_{1}} \\ 
m_{U_{2}} \\ 
m_{U_{3}}%
\end{array}%
\right) =\left( 
\begin{array}{c}
0.0011 \\ 
0.69 \\ 
173 \\ 
1313 \\ 
1507 \\ 
2261%
\end{array}%
\right) .  \label{mass of the quars}
\end{equation}%
It is clear that the light masses agree with the light masses of the SM \cite%
{Xing:2007fb, Xing:2011aa}. From the diagonalization of $\mathcal{M}_{d}$
and $\mathcal{M}_{u}$ we also obtain the non unitary $6\times 6$ CKM matrix $%
V$. Its moduli are given by 
\begin{equation}
\left\vert V\right\vert =\left( 
\begin{array}{cccccc}
0.97446 & 0.22459 & 0.003631 & 2.2\cdot 10^{-6} & 9.8\cdot 10^{-6} & 
2.9\cdot 10^{-5} \\ 
0.22446 & 0.97361 & 0.041118 & 2.8\cdot 10^{-5} & 2.3\cdot 10^{-4} & 
3.7\cdot 10^{-5} \\ 
0.00850 & 0.039901 & 0.987685 & 1.4\cdot 10^{-5} & 4.9\cdot 10^{-4} & 
3.7\cdot 10^{-4} \\ 
1.3\cdot 10^{-3} & 6.1\cdot 10^{-3} & 0.150913 & 2.1\cdot 10^{-6} & 7.5\cdot
10^{-5} & 5.7\cdot 10^{-5} \\ 
5.4\cdot 10^{-4} & 2.4\cdot 10^{-3} & 1.0\cdot 10^{-4} & 6.7\cdot 10^{-8} & 
5.5\cdot 10^{-7} & 9.0\cdot 10^{-8} \\ 
8.5\cdot 10^{-6} & 3.3\cdot 10^{-5} & 1.4\cdot 10^{-6} & 9.2\cdot 10^{-10} & 
7.6\cdot 10^{-9} & 1.2\cdot 10^{-9}%
\end{array}%
\right) ,  \label{Moduli VCKM}
\end{equation}%
and the arguments of its matrix elements are given in leading order by 
\begin{equation}
\arg \left( V\right) =\left( 
\begin{array}{cccccc}
0 & 6.4\cdot 10^{-4} & -1.197 & -8.2\cdot 10^{-3} & -3.109 & -2.3\cdot
10^{-3} \\ 
\pi & 0 & 0 & 1.22 & 0.668 & 0.666 \\ 
-0.393 & \pi +0.0188 & 0 & -1.93 & -2.54 & -2.13 \\ 
1.63 & -1.10 & 2.02 & 0.091 & -0.516 & -0.107 \\ 
1.57 & -1.56 & -1.56 & -0.345 & -0.895 & -0.896 \\ 
-1.09 & 1.55 & 1.57 & 2.79 & 2.25 & 2.17%
\end{array}%
\right) .  \label{Phases VCKM}
\end{equation}%
This generalized non-unitary CKM matrix $V$ deserves several comments:

\begin{enumerate}
\item The first upper left $3\times 3$ block reproduces to a great extent
the SM CKM mixing matrix, including the phases.

\item It is remarkable that all the moduli - except $|V_{tb}|$- of the $%
3\times 3$ light sector of $V$ agree with the SM fitted values within $%
1.5\sigma $, in fact most are within $1\sigma $.

\item A very important difference is in the element $|V_{tb}|$ that is
incompatible with the SM value. Disentangling this value from the SM one is
certainly an experimental challenge for single top production.

\item Looking at the four independent phases that can be defined in the $%
3\times 3$ light sector $\beta ,\gamma ,\beta _{s}$ and $\chi ^{\prime }$ -
related to the phases of $V_{td},V_{ub},V_{ts}$ and $V_{us}$, see references 
\cite{Branco:1999fs, AguilarSaavedra:2004mt} 
- we cannot find any relevant difference among the
phases of this example and the phases of the SM. We get in our model: 
\begin{eqnarray}
\beta =0.393, \ \gamma =1.197, \ \beta _{s} = 0.0188, \ \chi ^{\prime
}=0.000636
\end{eqnarray}

\item Since the new quark singlets do not couple directly to the $SU\left(
2\right) _{L}$ gauge bosons, in the limit of no mixing among chiral and
vectorial singlet quarks only the upper left $3\times 3$ sector of $V$ 
is different from zero. This explains the
smallness of the entries in the other sectors of $V$ and indicates that the
elements $V_{U_{i}d},V_{U_{i}s}$ and $V_{U_{i}b}$ can induce new physics
effects that could appear for example in $b\rightarrow d,b\rightarrow s$ or $%
s\rightarrow d$ transitions. Transitions induced by $V_{uD_{i}}$,$V_{cD_{i}}$
and $V_{tD_{i}}$ should be smaller than the latter.
\end{enumerate}

In summary, the CKM sector of our mixing matrix $V$ reproduces very well the
SM case except for a minor but definite deviation in $V_{tb}$ at the $1\%$
level.

\subsection{The FCNC structure}

In spite of the great similarity between the $3\times 3$ sector of our
mixing matrix $V$ and the SM CKM matrix, we know that in this model we will
have FCNC at tree level both in the up and in the down sectors. Of course we
expect to have these FCNC highly suppressed, but to check these expectations
we present the matrices that control these FCNC both in the couplings to the 
$Z$ and to the Higgs $h$ bosons. The moduli of the matrix elements that
control the FCNC in the up sector are given by $W^{u}=VV^{\dag } $: 
\begin{equation}
|W^{u}| = \left( 
\begin{array}{llllll}
1 & 1.1\cdot 10^{-8} & 2.9\cdot 10^{-17} & 4.5\cdot 10^{-18} & 4.7\cdot
10^{-6} & 4.0\cdot 10^{-6} \\ 
1.1\cdot 10^{-8} & 1 & 2.2\cdot 10^{-17} & 5.7\cdot 10^{-17} & 2.4\cdot
10^{-3} & 3.4\cdot 10^{-5} \\ 
2.9\cdot 10^{-17} & 2.2\cdot 10^{-17} & 0.977 & 0.149 & 1.7\cdot 10^{-16} & 
2.2\cdot 10^{-18} \\ 
4.5\cdot 10^{-18} & 5.7\cdot 10^{-17} & 0.149 & 2.28\cdot 10^{-2} & 2.6\cdot
10^{-17} & 3.3\cdot 10^{-19} \\ 
4.7\cdot 10^{-6} & 2.4\cdot 10^{-3} & 1.7\cdot 10^{-16} & 2.6\cdot 10^{-17}
& 5.8\cdot 10^{-6} & 8.1\cdot 10^{-8} \\ 
4.0\cdot 10^{-6} & 3.4\cdot 10^{-5} & 2.2\cdot 10^{-18} & 3.3\cdot 10^{-19}
& 8.1\cdot 10^{-8} & 1.1\cdot 10^{-9}%
\end{array}
\right).
\end{equation}
In the light sector $\left\vert W_{uc}^{u}\right\vert =1.1\cdot 10^{-8}$ is
too small to be in conflict with $D^{0}-\overline{D}^{0}$ mixing or $D^{0}
\rightarrow \mu \overline{\mu }$. Values like $\left\vert
W_{qt}^{u}\right\vert \sim 10^{-17}$ make $t\rightarrow Zu,Zc$ extremely
suppressed. The reminiscent of $V_{tb}\neq 1$ is here $\left\vert
W_{tt}^{u}\right\vert =0.977$ another challenging deviation from the SM to
be checked more likely in loops. It is important to realize that the light $%
3\times 3$ sector resembles very much the identity matrix ${1\>\!\!\!\mathrm{%
I}}_{3\times 3}$ except for $\left\vert W_{tt}^{u}\right\vert $. Values like 
$\left\vert W_{U_{1}t}^{u}\right\vert =0.149$ will dictate the neutral
current dominant decay channels of the heavy quark $U_{1}$ of mass $%
m_{U_{1}}=1313$ GeV: $U_{1}\rightarrow tZ$ and $U_{1}\rightarrow th$. The
moduli of the matrix elements that control the FCNC in the down sector are
given by $W^{d}=V^{\dag } V$: 
\begin{equation}
|W^{d}| = \left( 
\begin{array}{cccccc}
1 & 1.4\cdot 10^{-8} & 3.0\cdot 10^{-8} & 6.0\cdot 10^{-16} & 6.3\cdot
10^{-5} & 2.3\cdot 10^{-5} \\ 
1.4\cdot 10^{-8} & 1 & 1.4\cdot 10^{-7} & 2.8\cdot 10^{-5} & 2.4\cdot 10^{-4}
& 5.6\cdot 10^{-5} \\ 
3.0\cdot 10^{-8} & 1.4\cdot 10^{-7} & 1 & 1.3\cdot 10^{-5} & 4.9\cdot 10^{-4}
& 3.7\cdot 10^{-4} \\ 
6.0\cdot 10^{-6} & 2.8\cdot 10^{-5} & 1.3\cdot 10^{-5} & 9.6\cdot 10^{-10} & 
1.3\cdot 10^{-8} & 6.3\cdot 10^{-9} \\ 
6.3\cdot 10^{-5} & 2.4\cdot 10^{-4} & 4.9\cdot 10^{-4} & 1.3\cdot 10^{-8} & 
3.0\cdot 10^{-7} & 1.9\cdot 10^{-7} \\ 
2.3\cdot 10^{-5} & 5.6\cdot 10^{-5} & 3.7\cdot 10^{-4} & 6.3\cdot 10^{-9} & 
1.9\cdot 10^{-7} & 1.4\cdot 10^{-7}%
\end{array}
\right).
\end{equation}
Owing to the fact that $W^{d} $ can induce $K^{0}-\overline{K}^{0},B_{d}^{0}-%
\overline{B}_{d}^{0}$ and $B_{s}^{0}- \overline{B}_{s}^{0}$ mixing at tree
level through $Z$ exchange, the off-diagonal elements $W_{ds}^{d},W_{db}^{d}$
and $W_{sb}^{d}$ cannot be too large. But as we can see they are well below
(two to three orders of magnitude) the values from previous analyses 
\cite{Barenboim:2000zz, Barenboim:2001fd} 
that avoid conflicting with meson mixing constraints. The
bottom line is that the $3\times 3$ light sector of $|W^{d}|$ is very well
approximated by ${1\>\!\!\!\mathrm{I}}_{3\times 3}$.

But this is not the end of the story. We have now an enlarged $6\times 6$
CKM matrix $V$ with elements connecting light and heavy quarks of order $%
10^{-3}$ to $\ 10^{-4}$. These matrix elements enter into the loops that
generate FCNC with heavy quarks running inside the loop and with several
Inami-Lim (IL) functions \cite{Inami:1980fz} growing with the square of the
heavy quark masses. So, a priori, one has to check that the product of these
heavy masses with these suppressed couplings does not spoil the great
success of the SM in FCNC processes. At this point it is worthwhile
recalling that we have fixed our model with vectorial quarks by demanding
that a particular texture mass structure, imposed by symmetries, should
reproduce the light quark masses and the dominant CKM mixing matrix. On the
other hand, the symmetry does not impose any constraint on the product of
heavy masses square and mixing between light and heavy fermions. Therefore
we should check the relevant constraints.

\subsection{Loop FCNC constraints}

In general the structure of FCNC at one loop level in this model can be
quite involved. But, as we have seen, the tree level flavour changing
coupling are very much suppressed. So let us comment on the different
processes where, as we will see, tree level FCNC contributions can be safety
neglected:

\begin{enumerate}
\item The contributions that go with $W^{u}_{ij}$ and $W^{d}_{ij}$ in FCNC
tree level contributions to $\Delta F=1$ processes like $B_{q}\rightarrow
\mu \overline{\mu }.$ These NP pieces are proportional to $W^{d}_{bq}$ to be
compared with $\alpha _{em}V_{tb}^{\ast }V_{tq}$ times an IL function of
order one coming from the SM piece. The transitions where these NP
contributions are the largest ones are rare kaon decays where we have $%
W^{d}_{ds}/\alpha _{em}V_{td}^{\ast }V_{ts}\lesssim 10^{-2}.$ So for example 
$K^{+}\rightarrow \pi ^{+}\nu \overline{\nu }$ or $K_{L}\rightarrow \mu 
\overline{\mu }$ are not affected by these tree level contributions. The
corresponding decays in the $B$ meson systems are even less affected.

\item The contributions that go with $\left( W^{u}_{ij} \right)^{2}$ and $%
\left(W^{d}_{ij}\right)^{2}$ in FCNC tree level contributions to $\Delta F=2 
$ processes like $D^{0}-\overline{D}^{0}$ or $K^{0}-\overline{K} ^{0}$, $%
B_{d}^{0}-\overline{B}_{d}^{0}$ and $B_{s}^{0}-\overline{B}_{s}^{0}$. In
these cases, we are neglecting $W^{d}_{ds}$ with respect to $\sqrt{\alpha
_{em}}V_{td}^{\ast }V_{ts}$ and similarly for the $B_d$, $B_s$ and $D$
neutral meson systems. The largest NP correction appears in the kaon case
and is of order $10^{-6}$ times the SM contribution.

\item When the tree level FCNC are small, Barenboim and Botella \cite%
{Barenboim:1997pf} showed that the leading NP contribution to meson mixing,
induced by this FCNC enters at order $\alpha _{em}W^{d}_{ds}V_{td}^{\ast
}V_{ts}$, to be compared with $\alpha _{em}\left( V_{td}^{\ast
}V_{ts}\right) ^{2}$. So, in our example, these contributions are at most of
order $10^{-4} $. Again, we can neglect these contributions.
\end{enumerate}

Taking into account the previous considerations, we can analyze all the $%
\Delta F=1,2$ processes neglecting the tree level FCNC effects.

\subsubsection{$\Delta F=2$ pure loop constraints}

The neutral meson mixing is therefore dominated by the box diagrams that
generalize the SM one, with all species of heavy quarks plus the top and the
charm quarks - or the bottom and the strange quarks for $D^{0}-\overline{D}%
^{0}$ mixing - running inside the loop\footnote{%
Note that in the SM, the Inami-Lim function that appears in the box runs for 
$c$ and $t$ quarks after using unitarity $3\times 3$. In our case we have
for the kaon system, for example, $\sum_{i=1}^{6}V_{id}V_{is}^{\ast }=\left(
V^{\dag }V\right) _{ds}= W^{d}_{ds}$, so there is an additional contribution
- to what we are considering in the main text - proportional to the quark $u$
contribution and to $W^{d}_{ds}$, and therefore negligible. So we can extend
the sum from $c,t$ to $c,t,U_{1},U_{2},U_{3}$.}. If we define as usual $%
\lambda _{qq^{\prime }}^{a}=V_{aq^{\prime }}^{\ast }V_{aq}$ - for mesons
with down quarks -, the dominant corrections to the mixing $\left(
M_{12}\right) _{qq^{\prime }}$ of the meson with quark content $\left(
qq^{\prime}\right) $, with respect to the SM box diagram with internal top
quarks, can be written as: 
\begin{eqnarray}
\frac{(M_{12})_{qq^\prime}} {(M_{12}^{SM})_{qq^\prime}} &\sim
&1+\sum_{i=1}^{3}\left( \frac{\lambda _{qq^{\prime }}^{U_{i}}}{\lambda
_{qq^{\prime }}^{t}}\right) ^{2}\frac{S\left( x_{U_{i}}\right) }{S\left(
x_{t}\right) }+2\sum_{i=1}^{3}\left( \frac{\lambda _{qq^{\prime }}^{U_{i}}}{%
\lambda _{qq^{\prime }}^{t}}\right) \frac{S\left( x_{U_{i}},x_{t}\right) }{%
S\left( x_{t}\right) }  \label{mixing corrections} \\
&&+2\sum_{i<j}^{3}\frac{\left( \lambda _{qq^{\prime }}^{U_{i}}\lambda
_{qq^{\prime }}^{U_{j}}\right) }{\left( \lambda _{qq^{\prime }}^{t}\right)
^{2}}\frac{S\left( x_{U_{i}},x_{U_{j}}\right) }{S\left( x_{t}\right) }+\cdot
\cdot \cdot  \notag
\end{eqnarray}
where $S\left( x_{t}\right) ,$ $S\left( x_{U_{i}},x_{t}\right) $ are the IL
functions defined as in \cite{Branco:1999fs} and $x_{t},x_{U_{i}}=\left(
m_{t}/M_{W}\right) ^{2}$, $\left( m_{U_{i}}/M_{W}\right) ^{2}.$ In our case
we have for the first and second corrections in the kaon system 
\begin{eqnarray}
\left| \frac{\lambda _{sd}^{U_{1}}}{\lambda _{sd}^{t}}\right| ^{2}\frac{
S\left( x_{U_{1}}\right) }{S\left( x_{t}\right) } =1.6\times 10^{-2}, \qquad
2\left| \frac{\lambda _{sd}^{U_{1}}}{\lambda _{sd}^{t}}\right| \frac{S\left(
x_{U_{1}},x_{t}\right) }{S\left( x_{t}\right) } =0.13.
\end{eqnarray}
Including the charm and the leading QCD corrections, and defining the
corrections with respect the SM, 
\begin{eqnarray}
\Delta \left( P^{0}\right) =\left\vert \left( \frac{M_{12}}{M_{12}^{SM}}
\right)_{P^{0}} \right\vert -1 , \qquad \delta \phi \left( P^{0}\right)
=\left( \frac{\arg \left( M_{12}\right) }{ \arg \left( M_{12}^{SM}\right) }%
\right) _{P^{0}}-1,  \label{Definitions Corrections Mixing}
\end{eqnarray}
we get for the $K^{0},B_{d}$ and $B_{s}$ mixings 
\begin{equation}
\left( 
\begin{array}{c}
\Delta \left( K^{0}\right) \\ 
\Delta \left( B_{d}^{0}\right) \\ 
\Delta \left( B_{s}^{0}\right)%
\end{array}
\right) =\left( 
\begin{array}{c}
0.02 \\ 
0.12 \\ 
0.12%
\end{array}
\right) \text{ },\text{ }\left( 
\begin{array}{c}
\delta \phi \left( K^{0}\right) \\ 
\delta \phi \left( B_{d}^{0}\right) \\ 
\delta \phi \left( B_{s}^{0}\right)%
\end{array}
\right) =\left( 
\begin{array}{c}
0.12 \\ 
0.01 \\ 
0.01%
\end{array}
\right).  \label{Corrections Mixing Model 1}
\end{equation}
We conclude that these kinds of models are compatible with the actual
analysis beyond the SM, but they can have sizeable effects. For example this
model does not modify $\Delta M_{B_{d}}/\Delta M_{B_{s}}$ but can give 12\%
corrections to $\Delta M_{B_{d}}$ and $\Delta M_{B_{s}}$. Furthermore, $%
\epsilon _{K}$ can have 12\% corrections.

\subsubsection{$\Delta F=1$ pure loop constraints}

In these processes - for example $q\rightarrow q^{\prime }\mu \overline{\mu }
$ - the corrections to the top dominated SM amplitudes (when necessary charm
has to be taken into account) are given by 
\begin{eqnarray}
\frac{A\left( q\rightarrow q^{\prime }\mu \overline{\mu }\right) }{A\left(
q\rightarrow q^{\prime }\mu \overline{\mu }\right) _{SM}} &\sim
&1+\sum_{i=1}^{3} \frac{\lambda _{qq^{\prime}}^{U_i} Y\left( x_{U_{i}}\right)%
}{\lambda _{qq^{\prime }}^{t} Y \left( x_t \right)}  \notag \\
&&-\sum_{i,j=c,t, U_{1},U_{2},U_{3}} \frac{V_{iq^{\prime }}^{\ast }\left(
W^{u}-I\right) _{ij}V_{jq}N\left( x_{i},x_{j}\right) }{\lambda _{qq^{\prime
}}^{t}Y\left( x_{t}\right) }  \label{delta F=1 corrections}
\end{eqnarray}
The IL functions $Y\left( x_{U_{i}}\right) $ grow with the square of the
quark mass $U_{i}$ - also $N\left( x_{U_{i}},x_{U_{i}}\right) $ -. This
fact, apparently, will make the second term more relevant than the
corresponding one in $\Delta F=2$ processes: here there is only a $\lambda
_{qq^{\prime }}^{U_{i}}/\lambda _{qq^{\prime }}^{t}$ suppression and not a $%
\left( \lambda _{qq^{\prime }}^{U_{i}}/\lambda _{qq^{\prime }}^{t}\right)
^{2}$ suppression as in Eq.~(\ref{mixing corrections}). But $%
W^{u}_{U_{i}U_{i}}$ are very small - the new quarks are singlets under $%
SU\left( 2\right) _{L}$-, therefore the last term has a piece enforcing
decoupling and cancelling partially the second piece \footnote{%
Note that this term is usually overlooked in the literature.}. For more
details one can see references \cite{Nardi:1995fq, Vysotsky:2006fx,
Kopnin:2008ca, Picek:2008dd, inprogress}. A similar structure appears in $%
q\rightarrow q^{\prime }\nu \overline{\nu }$. If we define, as in the $%
\Delta F=2$ processes, the deviation from the SM model 
\begin{equation*}
r\left( A\rightarrow B\right) =\frac{\Gamma \left( A\rightarrow B\right) }{
\Gamma \left( A\rightarrow B\right) _{SM}}-1,
\end{equation*}
we get 
\begin{equation}
\begin{array}{lll}
r\left( K_{L}\rightarrow \mu \overline{\mu }\right) _{SD}=0.32, & r\left(
B_{d}\rightarrow \mu \overline{\mu }\right) =0.31, & r\left(
B_{s}\rightarrow \mu \overline{\mu }\right) =0.30, \\ 
r\left( K^{+}\rightarrow \pi ^{+}\nu \overline{\nu }\right) =0.20, & r\left(
B^{+}\rightarrow \pi ^{+}\nu \overline{\nu }\right) =0.21, & r\left(
B^{+}\rightarrow K^{+}\nu \overline{\nu }\right) = 0.20.%
\end{array}
\label{Delta Fequal 1 processes model 1}
\end{equation}
Some of these predictions can be definitely excluded or verified very soon
by the LHC experiments.

We also have analyzed other loop mediated processes and, for example, we get
for the oblique corrections \cite{Peskin:1991sw, Lavoura:1992np}: 
\begin{equation}
\Delta T=0.22\text{ },\text{ }\Delta S=0.06, \ \ \Delta U=0.003.
\end{equation}

\subsection{The heavy vector-like quark decay channels}

In our model the new heavy quarks decay mainly throughout charged or neutral
currents. The decays can be characterized by $Q_{j}\rightarrow q_{i}B.$
where $Q_{j}=U_{j},D_{j}$, is the new heavy fermion, $B=Z,h,W^{\pm}$ and $%
q_{i}$ the final state fermion, namely $u,c,t,d,s$ or $b$. The different
partial decay widths can be written as 
\begin{equation}
\Gamma \left( Q_{j}\rightarrow q_{i}B\right) =\xi _{B}\frac{m_{Q_{j}}^{2}}{%
32\pi \upsilon ^{2}}\left\vert X_{ij}^{B,Q_{j}}\right\vert ^{2}f_{B}\left(
x_{j}^{B},r_{ij}\right)
\end{equation}
where 
\begin{equation}
x_{j}^{B}=\frac{M_{B}^{2}}{m_{Q_{j}}^{2}},\text{ }r_{ij}=\frac{ m_{q_{i}}^{2}%
}{m_{Q_{j}}^{2}},\text{ } \xi _{ Z } = \xi _{ h } = 1, \ \xi _{ W^{\pm} } =
2,  \notag
\end{equation}
\begin{equation}
\mbox{and} \ X_{ij}^{B,Q_{j}}=\left\{ 
\begin{array}{c}
V_{ij} \ \mbox{for} \ B=W^- , \\ 
V_{ji} \ \mbox{for} \ B=W^+, \\ 
W_{ij}^{Q} \ \mbox{for} \ B= h, Z, \\ 
\end{array}
\right.
\end{equation}
and finally, 
\begin{eqnarray}
f_{W^{\pm }}\left( x,r\right) = f_{Z}\left( x,r\right), \qquad f_{h}\left(
x,r\right) \sim f_{Z}\left( x,r\right) \sim 1,
\end{eqnarray}
where the last two relations are valid when $m_{q_{i}},M_{Z},M_{W},M_{h}\ll
m_{Q_{j}}$. Detailed formulas are included in Appendix B. In this regime, to
a very good approximation \cite{delAguila:1989rq}, 
\begin{equation}
\Gamma \left( Q_{j}\rightarrow q_{i}Z\right) \simeq \Gamma \left(
Q_{j}\rightarrow q_{i}h\right).
\end{equation}
Furthermore, under the same conditions,we also have 
\begin{eqnarray}
\Gamma \left( U_{j}\rightarrow d_{i}^{\prime }W\right) &\simeq &2\frac{
\left\vert V_{ji}\right\vert ^{2}}{\left\vert W_{ij}^{u}\right\vert ^{2}}
\Gamma \left( U_{j}\rightarrow u_{i}Z\right),  \notag \\
\Gamma \left( D_{j}\rightarrow u_{i}^{\prime }W\right) &\simeq &2\frac{
\left\vert V_{ij}\right\vert ^{2}}{\left\vert W_{ij}^{d}\right\vert ^{2}}
\Gamma \left( D_{j}\rightarrow d_{i}Z\right),
\end{eqnarray}
for $q_{i}$ and $q_{i}^{\prime }$ in the same generation, as suggested by
the subindex $i$: for example $c$ and $s$. It turns out that, under
reasonable conditions explained in Appendix B, it is quite common to have $%
\left\vert V_{ji}\right\vert ^{2}\sim \left\vert W_{ij}^{u}\right\vert ^{2}$
and $\left\vert V_{ij}\right\vert ^{2}\sim \left\vert W_{ij}^{d}\right\vert
^{2}$. Therefore we also have 
\begin{equation}
\Gamma \left( Q_{j}\rightarrow q_{i}^{\prime }W\right) \simeq 2\Gamma \left(
Q_{j}\rightarrow q_{i}Z\right).
\end{equation}
Although many searches for new heavy quarks assume $\Gamma \left(
Q_{j}\rightarrow q_{i}Z\right) :\Gamma \left( Q_{j}\rightarrow q_{i}h\right)
:\Gamma \left( Q_{j}\rightarrow q_{i}^{\prime }W\right) =1:1:2$ and $%
B_{r}\left( Q_{j}\rightarrow q_{i}Z\right) +B_{r}\left( Q_{j}\rightarrow
q_{i}h\right) +B_{r}\left( Q_{j}\rightarrow q_{i}^{\prime }W\right) \simeq 1$%
, the later is unjustified and the total $Q_{j}$ decay width can be
distributed among the decay channels to different generations while the 1 :
1 : 2 pattern of branching ratios `` per generation" is maintained \cite{Barger:1995dd}. This
fact happens in the present model and we show in Tables 3 and 4 the dominant
decay channels of the heavy quarks. 
\begin{table}[tbp]
\centering
\begin{tabular}{|c|c||c|c|c||c|c|c||c|c|c|}
\cline{2-11}
\multicolumn{1}{c|}{} & \multirow{2}{*}{Width (MeV)} & \multicolumn{9}{|c|}{
Branching ratio to channel (\%):} \\ \cline{3-11}
\multicolumn{1}{c|}{} &  & $Zd$ & $Zs$ & $Zb$ & $hd$ & $hs$ & $hb$ & $Wu$ & $%
Wc$ & $Wt$ \\ \hline
$D_{1}$ & $2.9\cdot 10^{-4}$ & 0.9 & 20.3 & 4.4 & 0.9 & 19.3 & 4.2 & 0.2 & 
40.7 & 8.8 \\ \hline
$D_{2}$ & 0.81 & 0.3 & 4.9 & 20.4 & 0.3 & 4.9 & 20.2 & 0 & 8.9 & 40.0 \\ 
\hline
$D_{3}$ & 0.69 & 0 & 0.5 & 24.9 & 0 & 0.5 & 24.7 & 0.3 & 0.5 & 48.1 \\ \hline
\end{tabular}%
\caption{Decays of new down type quaks}
\end{table}

\begin{table}[tbp]
\centering
\begin{tabular}{|c|c||c|c|c||c|c|c||c|c|c|}
\cline{2-11}
\multicolumn{1}{c|}{} & \multirow{2}{*}{Width (GeV)} & \multicolumn{9}{|c|}{
Branching ratio to channel (\%):} \\ \cline{3-11}
\multicolumn{1}{c|}{} &  & $Zu$ & $Zc$ & $Zt$ & $hu$ & $hc$ & $ht$ & $Wd$ & $%
Ws$ & $Wb$ \\ \hline
$U_{1}$ & 32.9 & 0 & 0 & 23.9 & 0 & 0 & 24.7 & 0 & 0 & 51.4 \\ \hline
$U_{2}$ & $1.3\cdot 10^{-2}$ & 0 & 25.1 & 0 & 0 & 24.7 & 0 & 2.5 & 47.6 & 0
\\ \hline
$U_{3}$ & $8.6\cdot 10^{-6}$ & 0.3 & 24.7 & 0 & 0.3 & 24.5 & 0 & 3.2 & 46.8
& 0 \\ \hline
\end{tabular}%
\caption{Decays of new up type quaks}
\end{table}

For example, $D_{1}$ - the lightest of the heavy vector-like quarks in this
model - has around $20\%$ of its branching ratio to decay channels in the
third generation and some $80\%$ decaying to the second generation $\left(
Zs,hs,Wc\right)$. These are not the most common ways of searching for this
down type vector-like quarks although the different results by ATLAS and CMS
can be adapted. Note that neither the decay to the third family is the
dominant one nor there is a unique family entering in the dominant decay
products. Similar patterns appear in $D_{2}$ decays. As far as the up
vector-like quarks are concerned the lightest $U_{1}$ decays dominantly to
the third family while the other (heavier) two decay to the second family.

\subsection{Other examples}

We have presented a paradigmatic example to show that it is possible and
even relatively easy to accomplish our goal. Indeed, in our example the
light quark masses and CKM mixing are generated from the couplings of
ordinary quarks to the new heavy vector-like quarks, keeping the extended
symmetry that explains why to first order the CKM matrix is the identity. In
addition, we predict in the flavour electroweak phenomenology many
deviations from the SM, some of them at 2$\sigma $ level or more. Note that
we have also a definite deviation of $3\times 3$ unitarity, albeit very
difficult to disentangle experimentally: $|V_{tb}|\sim 0.9877$. Also we have
six new heavy quarks that can be seen at LHC, some of them with very
peculiar characteristic decays. From the present example, one can get the
wrong impression that if these new quarks are not discovered at the LHC and
if the different deviations predicted in the flavour and electroweak sectors
are not established, the main ideas of this paper are no longer relevant.
This is not the case. On the contrary, we will show that there are many
additional solutions which, without spoiling the main characteristics of the
low mass sector, have the feature that New Physics effects smoothly decouple
in the low energy phenomenology. In order to see how this decoupling arises,
let us consider Eq.~(\ref{not}) and label our explicit example in Eqs.~(\ref%
{masas pequenas}) to (\ref{h_l singlet mass matrix 2}) with a superscript
``(1)", i.e., the matrices in Eqs.~(\ref{masas pequenas}) to (\ref{h_l
singlet mass matrix 2}) are 
\begin{equation}
\mathcal{M}_{d}^{\left( 1\right) }=\left( 
\begin{array}{cc}
m_{d} & \omega _{d} \\ 
X_{d}^{\left( 1\right) } & M_{d}^{\left( 1\right) }%
\end{array}
\right), \text{ \ }\mathcal{M}_{u}^{\left( 1\right) }=\left( 
\begin{array}{cc}
m_{u} & \omega _{u} \\ 
X_{u}^{\left( 1\right) } & M_{u}^{\left( 1\right) }%
\end{array}
\right).
\end{equation}
Let us construct a second solution, labelled ``(2)" that differ by the real
numbers $\rho _{d}$ and $\rho _{u}$: 
\begin{equation}
\mathcal{M}_{d}^{\left( 2\right) }=\left( 
\begin{array}{cc}
m_{d} & \omega _{d} \\ 
\rho _{d}X_{d}^{\left( 1\right) } & \rho _{d}M_{d}^{\left( 1\right) }%
\end{array}
\right), \text{ \ }\mathcal{M}_{u}^{\left( 2\right) }=\left( 
\begin{array}{cc}
m_{u} & \omega _{u} \\ 
\rho _{u}X_{u}^{\left( 1\right) } & \rho _{u}M_{u}^{\left( 1\right) }%
\end{array}
\right).  \label{54}
\end{equation}
Now, for $\rho _{d}, \rho _{u} > 1$ we obtain another solution which
reproduces ``essentially" the same light quark masses and mixings, while
deviations from the SM decouple as $\rho _{d}$ and $\rho _{u}$ are
increased. The basic equations are Eqs.~(\ref{23}) to (\ref{28}). Equation (%
\ref{28}) shows that $\mathcal{H}_{eff}^u$ does not change at leading order
in going from $\mathcal{M} _{u}^{\left( 1\right) }$ to $\mathcal{M}%
_{u}^{\left( 2\right) }$ and so $K_{u}^{\left( 2\right) }\sim K_{u}^{\left(
1\right) }$. At the same time Eqs.~(\ref{24}) to (\ref{26}) tell us that $%
R_{u}^{\left( 1\right) }$ and $S_{u}^{\left( 1\right) }$ scale as $%
R_{u}^{\left( 2\right) }\sim \rho _{u}^{-1}R_{u}^{\left( 1\right) }$ and $%
S_{u}^{\left( 2\right) }\sim \rho _{u}^{-1}S_{u}^{\left( 1\right) }$ which
leads to a small change in $K_{u}^{\left( 2\right) }$ taking it closer to
unitarity. The invariance of $\mathcal{H} _{eff}^{u \left( 1\right) }
\rightarrow \mathcal{H} _{eff}^{u \left( 2\right) } $ under the scaling in
Eq.~(\ref{54}) is at the origin of the fact that, once we have a solution of
the type presented here, we have a continuos of solutions with essentially
the same light quark masses and mixing and with heavy quarks much heavier
and more decoupled. It remains to check that the effects from heavy quarks
running inside the loops also decouple. This can be easily seen by realizing
that our CKM matrix can also be written: 
\begin{equation}
V=\left( 
\begin{array}{c}
K_{u}^{\dag } \\ 
R_{u}^{\dag }%
\end{array}
\right) \left( 
\begin{array}{cc}
K_{d} & R_{d}%
\end{array}
\right) =\left( 
\begin{array}{cc}
K_{u}^{\dag }K_{d} & K_{u}^{\dag }R_{d} \\ 
R_{u}^{\dag }K_{d} & R_{u}^{\dag }R_{d}%
\end{array}
\right)
\end{equation}
From here it is evident that the submatrix $R_{u}^{\dag }K_{d}$ scales to $%
\rho _{u}^{-1}R_{u}^{\dag }K_{d}$ and consequently we have a very simple
scaling law from the up heavy sector scaling: 
\begin{equation}
\begin{array}{ccc}
X_{u}\rightarrow \rho _{u}X_{u} & , & M_{u}\rightarrow \rho _{u}M_{u}, \\ 
V_{U_{i}d_{j}}\rightarrow \rho _{u}^{-1}V_{U_{i}d_{j}} & , & \lambda
_{d_{i}d_{j}}^{U_{i}}\rightarrow \rho _{u} ^{-2}\lambda
_{d_{i}d_{j}}^{U_{i}}, \\ 
m_{U_{i}}\rightarrow \rho _{u}m_{U_{i}}. &  & 
\end{array}%
\end{equation}
This scaling is sufficient - if needed - to suppress all loop induced
effects. For example, in the case of $\Delta F=2$ processes, the first
correction in Eq.~(\ref{mixing corrections}) scales as $\left( \lambda
_{qq^{\prime }}^{U_{i}}\right) ^{2}S\left( x_{U_{i}}\right) \sim \left(
\lambda _{qq^{\prime }}^{U_{i}}\right) ^{2}m_{U_{i}}^{2}$ and therefore it
decreases as $\rho _{u}^{-2}$. The second correction scale as $\rho
_{u}^{-2}\ln \rho _{u} $ according to the behaviour of the IL function $%
S\left( x_{U_{i}},x_{t}\right) $. Behaviour similar to the last one appear
in the $\Delta F=1$ processes following the cancellations explained in Eq.~(%
\ref{delta F=1 corrections}).

To be more specific we present briefly the results for a second example with 
$\mathcal{M}_{u}^{\left( 2\right) }$ constructed with $\rho _{u}=2$ and
keeping the rest of the model unchanged. For $\Delta F=2$ processes\ we get,
instead of the results in Eq.~(\ref{Corrections Mixing Model 1}), the
following 
\begin{equation}
\left( 
\begin{array}{c}
\Delta \left( K^{0}\right) \\ 
\Delta \left( B_{d}^{0}\right) \\ 
\Delta \left( B_{s}^{0}\right)%
\end{array}%
\right) =\left( 
\begin{array}{c}
0.01 \\ 
0.07 \\ 
0.06%
\end{array}
\right), \ \ \left( 
\begin{array}{c}
\delta \phi \left( K^{0}\right) \\ 
\delta \phi \left( B_{d}^{0}\right) \\ 
\delta \phi \left( B_{s}^{0}\right)%
\end{array}
\right) =\left( 
\begin{array}{c}
0.06 \\ 
0.01 \\ 
0.01%
\end{array}
\right).
\end{equation}
On average, the corrections to the SM mixing values get reduced from a 12\%
to some 6\%. In the case of $\Delta F=1$ processes instead of Eq.~(\ref%
{Delta Fequal 1 processes model 1}) we get 
\begin{equation}
\begin{array}{lll}
r\left( K_{L}\rightarrow \mu \overline{\mu }\right) _{SD}=0.16, & r\left(
B_{d}\rightarrow \mu \overline{\mu }\right) =0.16, & r\left(
B_{s}\rightarrow \mu \overline{\mu }\right) =0.16, \\ 
r\left( K^{+}\rightarrow \pi ^{+}\nu \overline{\nu }\right) =0.11, & r\left(
B^{+}\rightarrow \pi ^{+}\nu \overline{\nu }\right) =0.11, & r\left(
B^{+}\rightarrow K^{+}\nu \overline{\nu }\right) = 0.10.%
\end{array}
\label{58}
\end{equation}
In this sector, the deviation from the SM model gets reduced by a factor of
2. In this model ``(2)" the deviations in these processes are at the 11\% to
16\% level, showing a smooth decoupling of the effects while at the same
time the heavy up-type quarks get heavier masses: $\left(
m_{U_{1}},m_{U_{2}},m_{U3}\right) \sim \left( 2.6,3,4.5\right) $ TeV. As for
the oblique corrections, they have the following values: 
\begin{equation}
\Delta T=0.11, \ \ \Delta S=0.05, \ \ \Delta U=0.004.
\end{equation}
Finally for $\left( |V_{td}|, |V_{ts}|, |V_{tb}|\right) $ we get $\left(
0.008575,0.040258,0.99632\right) $ since $K_{u}$ deviates less from
unitarity as consequence of the reduction of $R_{u}^{\dag }K_{d}$ by a
factor of 2.

\section{Conclusions}

We have presented a simple solution to a novel fine-tuning problem present
in the SM. The solution involves the introduction of a flavour symmetry G,
together with vector-like quarks of $Q=-1/3$ and $Q=2/3$ charges, as well as
a complex singlet scalar. In the absence of vector-like quarks only the
bottom and the top quarks acquire mass and $V_{CKM}={1\>\!\!\!\mathrm{I}}$.
We have shown that in the presence of the vector-like quarks which mix with
the standard quarks, a realistic quark mass spectrum can be obtained and a
correct CKM matrix can generated. It is remarkable that these results are
obtained in a framework where G is an exact symmetry of the Lagrangian, only
spontaneously broken by the vacuum. We have presented specific realistic
examples and have analysed various FCNC processes as well as the decay
channels of the vector-like quarks. It is also remarkable that in the
framework of fully realistic models, some of these vector-like quarks are at
the reach of the second LHC run.

\section*{Acknowledgments}

The authors acknowledge financial support from the Spanish MINECO under
grant FPA2015-68318-R, by the Severo Ochoa Excellence Center Project
SEV-2014-0398, by Generalitat Valenciana under grant PROMETEOII/ 2014/049
and by Funda\c{c}\~ao para a Ci\^encia e a Tecnologia (FCT, Portugal)
through the projects CERN/FIS-NUC/0010/2015, and CFTP-FCT Unit 777
(UID/FIS/00777/2013) which are partially funded through POCTI (FEDER),
COMPETE, QREN and EU. M.N. acknowledges support from FCT through the
postdoctoral grant SFRH/BPD/112999/2015. The authors also acknowledge the
hospitality of Universidad de Valencia, IFIC, and CFTP at IST Lisboa during
visits for scientific collaboration.

\section*{Appendix A. Derivation of effective Hermitian squared mass matrix}

In order to derive the expression given for $\mathcal{H}_{eff}$ in Eq.~(\ref%
{28}) we start from: 
\begin{equation}
\mathcal{M}\mathcal{M}^\dagger \mathcal{U}_{L} = \mathcal{U}_{L} \left( 
\begin{array}{cc}
d^2 & o \\ 
0 & D^2%
\end{array}
\right)
\end{equation}
using the notation of Eq.~(\ref{plus}), where for simplicity, we have
dropped the indices $d$ and $u$ for down and up. We can rewrite this
equation explicitly, in terms of each one of the four $3 \times 3$ blocks,
using Eq.~(\ref{not}) for the matrix $\mathcal{M}$: 
\begin{eqnarray}
(m m^\dagger + \omega \omega^\dagger)\ K + (mX^\dagger + \omega M^\dagger)\
S = K \ d^2,  \label{eqa} \\
(m m^\dagger + \omega \omega^\dagger)\ R + (mX^\dagger + \omega M^\dagger)\
T = R \ D^2, \\
(X m^\dagger + M \omega^\dagger)\ K + (XX^\dagger + M M^\dagger)\ S = S \
d^2,  \label{eqc} \\
(X m^\dagger + M \omega^\dagger)\ R + (XX^\dagger + M M^\dagger)\ T = T \
D^2.
\end{eqnarray}
Since the term $S\ d^2$ is much smaller than the other two, Eq.~(\ref{eqc})
can be approximated by: 
\begin{equation}
(XX^\dagger + M M^\dagger)\ S \simeq - (X m^\dagger + M \omega^\dagger)\ K
\end{equation}
or else 
\begin{equation}
S \simeq - (XX^\dagger + M M^\dagger)^{-1} (X m^\dagger + M \omega^\dagger)\
K
\end{equation}
Replacing $S$ in Eq.~(\ref{eqa}) we obtain to a good approximation: 
\begin{equation}
[ (m m^\dagger + \omega \omega^\dagger) - (mX^\dagger + \omega M^\dagger)
(XX^\dagger + M M^\dagger)^{-1} (X m^\dagger + M \omega^\dagger)]\ K = K \
d^2
\end{equation}
implying: 
\begin{equation}
K^{-1} \mathcal{H}_{eff} K = d^2
\end{equation}
with $\mathcal{H}_{eff}$ given by Eq.~(\ref{28}).

\section*{Appendix B. Vector-like quark decays}

The functions that appear in section 4.3 that are needed for the heavy quark
decays are 
\begin{align}
f_h (x, y) & = (1+y-x) f(x, y),  \notag \\
f_V (x, y) & = \left( (1-y^2) + x(1+y) -2 x^2 \right) f(x, y), \qquad V= Z,
W^{\pm},
\end{align}
where 
\begin{equation}
f(x, y) = \sqrt{1 - \left( \sqrt{y} + \sqrt{x}\right)^2} \sqrt{1 - \left( 
\sqrt{y} - \sqrt{x}\right)^2}
\end{equation}
Let us explain the origin and validity of the relations 
\begin{equation}
|V_{ji^\prime}|^2 \sim |(W^u)_{ij}|^2, \qquad |V_{i^\prime j}|^2 \sim
|(W^d)_{ij}|^2,  \label{eq62}
\end{equation}
where the prime means that $i$ and $i^\prime$ correspond to the same
generation number, $i= i^\prime$. For that purpose we introduce, in a self
explanatory matrix notation, 
\begin{equation}
V = \left( 
\begin{array}{cc}
V_{ud} & V_{uD} \\ 
V_{Ud} & V_{UD}
\label{self}
\end{array}
\right),
\end{equation}
where each submatrix connects the corresponding types of quarks $u,U$ and $%
d, D$ (in the present model, all four submatrices are $3 \times 3$).
Similarly, we also introduce 
\begin{equation}
W^u = \left( 
\begin{array}{cc}
W_{uu}^u & W_{uU}^u \\ 
W_{Uu}^u & W_{UU}^u%
\end{array}
\right), \qquad W^d = \left( 
\begin{array}{cc}
W_{dd}^d & W_{dD}^d \\ 
W_{Dd}^d & W_{DD}^d%
\end{array}
\right).
\end{equation}
Following again the notation in section 3.3 for the up and down sectors, we
have 
\begin{equation}
V_{Ud} = R_u^\dagger K_d, \qquad W_{Uu}^u = R_u^\dagger K_u,
\end{equation}
so we can write 
\begin{equation}
V_{Ud} = W_{Uu}^u K_u^{-1} K_d
\end{equation}
At leading order $K_u^{-1} \sim K_u^\dagger$ and we get, to a high accuracy
(and similarly for $V_{uD}$ and $W_{dD}^d$) 
\begin{equation}
V_{Ud} \sim W_{Uu}^u V_{ud}, \qquad V_{uD} \sim V_{ud} W_{dD}^d
\end{equation}
From these expressions it is easy to prove that relations in Eq.~(\ref{eq62}%
) hold, at least for the dominant decay channel of the corresponding heavy
up or down quark. The key ingredient is that $V_{ud} \sim {1\>\!\!\!\mathrm{I%
}} $, in Eq.~(\ref{self}), with corrections at most of order $\lambda = 0.22$.

\end{document}